\documentstyle[preprint,aps]{revtex} 
\begin{document}
\draft
\title{\bf {Conductance Quantization in a Periodically
Modulated Quantum Channel: backscattering and mode mixing}}
\author{P. Singha Deo$^1$ \cite{eml1}, B. C. Gupta$^2$, A. M.
Jayannavar$^2$, and F. M. Peeters$^{1,}$ \cite{eml2}}
\address{$^1$ Dept. of Physics, University of Antwerp (UIA),
Universiteitsplein 1,\\ B-2610 Antwerpen, Belgium.\\
$^2$ Institute of Physics, Bhubaneswar 751005, India}
\maketitle
\begin{abstract}
It is known that the conductance of a quantum point contact is
quantized in units of $2e^2/h$ and this quantization is destroyed by
a non-adiabatic scatterer in the point contact, due to
backscattering.  Recently, it was shown [Phys. Rev. Lett. {\bf 71},
137 (1993)] that taking many non-adiabatic scatterers periodically in
a quantum channel, the quantization can be recovered. We study this
conductance quantization of a periodic system in the presence of a
strong defect. A periodic arrangement of double-stubs give remarkable
quantization of conductance. A periodic arrangement of
double-constrictions also gives a very good quantization only when
the separation between the constrictions is small. We conclude that
conductance quantization of a periodically modulated channel is
robust.

\end{abstract}
\narrowtext

\section{Introduction}

In a novel experiment van Wees {\it et al} \cite{wee} and Wharam {\it
et al} \cite{wha} found that the conductance of a nano-meter scale
constriction between two 2D electron gas increases monotonically in
steps of 2$e^2/h$ as more and more propagating sub-bands become
available. Such a constriction is referred to as a quantum point
contact (QPC). This quantization of conductance is due to the lateral
quantization of the energy (due to the formation of sub-bands) and
sheds light on the physical relevance of Landauer's formula
\cite{lan}.  Transport along the wire with such quantized states
become reflection-less due to the absence of backscattering and then
the conductance quantization is simply evident from the Landauer
conductance formula.  Energy can also be quantized by applying a
large magnetic field. In such a situation the four probe Hall
conductance is found to be quantized and the phenomenon is well known
as the Integral Quantum Hall Effect (IQHE) \cite{tap}.  Whereas in
IQHE the quantization has been observed to an accuracy of one part in
billion, the accuracy in the QPC experiments is one part in hundred.
This departure was therefore taken very seriously.  Since then the
conductance of QPC has been studied in great details, experimentally
as well as theoretically.  Various aspects like the role of point
defects in the point contact (PC) \cite{bag}, the role of evanescent
modes in the PC \cite{bag,ber}, mode matching at the junction of the
contact and the reservoir \cite{kir}, adiabatic as well as
non-adiabatic shape changes of the point contact \cite{sza,hua},
non-uniformity of potential in th PC \cite{cir}, bound states in the
presence of non-adiabatic shape changes \cite{pee,nix,but}, periodic
modulations in the point contact \cite{hua,kou,bru,aki}, magnetic
field in the PC \cite{noc}, etc. have been studied in details.
Non-adiabatic constrictions in the point contact roughly maintain the
quantization \cite{sza}, but if there is a single non-adiabatic
cavity in the point contact then this quantization is completely
washed away \cite{hua}.  Recently Leng and Lent \cite{len} have shown
that the quantization of the conductance is recovered if a large
number of non-adiabatic constrictions are taken in the point contact
periodically, however, the conductance does not increase
monotonically with the energy. It was analyzed by Leng and Lent
\cite{len} that back scattering can be suppressed in these
periodically modulated channels (except in the narrow regions between
resonances which can not be avoided in a finite periodic system)
because the periodicity results in Bloch state like good quantum
states and thus resulting in a perfect conductance quantization.
However, the periodic system has to be connected to leads on both
sides, and due to matching of wave-function between the leads and the
periodic system there will be some mixing of the transverse modes
(modes arising due to the finite thickness in the direction
perpendicular to the propagating direction). In the presence of some
channel mixing it is possible that all the resonances in the
transmission probabilities are not due to the symmetry of the system
but some of them are Fano resonances, arising due to the mixing of
transverse modes \cite{bag}. Hence, if there are defects or disorder
in the system, which will result in the breakdown of the symmetry of
the system, it is possible that this may not affect the Fano
resonances much, although it will obviously affect the resonances
arising due to the symmetry. Hence, it may be interesting to see the
effect of a defect on the quantized conductance of a periodically
modulated channel.

Earlier we have reported \cite{pro} the effect of a defect on the
conductance of a periodically modulated channel when only one
propagating sub-band is populated in the main wire and we showed that
the defect has very drastic effects on the conductance which can be
used for efficient band-tailoring. In this paper we study the effect
of similar defects on the conductance when two or more sub-bands are
occupied.

This paper is organized as follows. In section II we outline the
numerical technique used. In section III we graphically discuss our
results. In section IV our conclusions are summarized. And finally
section V is devoted to acknowledgments.

\section{Theoretical treatment}

Fig.~1 shows a single non-adiabatic geometric scatterer in a point
contact.  The point contact is bounded between $-\frac{b}{2}<$ x
$<\frac{b} {2}$ whereas the scatterer in the point contact is bounded
between $-\frac{c}{2}<$ y $<\frac{c}{2}$. The width of the scatterer
in the $x$ direction is $2a$. When $c > b$ we refer to the system as
a double-stub whereas for $c < b$ we refer to the system as a
double-constriction. The scatterer is symmetric in the $y$ direction.
So the sub-bands will have a definite parity and opposite parity
states can never mix.  By cascading many such units we can make a
periodically modulated QPC or quantum channel. Reservoirs are
attached to two ends of the sample. We do not consider any mode
matching at the junctions of the reservoirs.  As a result we do not
have evanescent modes in the main wire but we do have sufficient
number of evanescent modes inside the non-adiabatic scatterer. In a
physical situation where the leads connecting the sample to the
reservoirs become large, the effect of this mode matching becomes
unimportant \cite{aki}.  And moreover, if we include evanescent modes
in the main wire, the system of periodic arrangement of double stubs
and the system of periodic arrangement of double-constrictions become
similar to each other.  We deliberately make them different to give
us some physical insight.

We have evaluated the conductance or the transmission coefficient
across the system numerically and the procedure followed is briefly
outlined below. The details can be found in Ref. \cite{bay}. The
system is unbounded in the propagating direction, {\em i.e.}, the $x$
direction whereas in the $y$ direction the system is bounded due to
the finite width. The hard wall boundary conditions are assumed along
the boundaries of the channel and these conditions define the channel
modulation.  Inside the system we have taken the magnitude of the
quantum mechanical potential $V$=0, {\em i.e.}, we are in the
classically unbound regime. We first consider the case of a single
scatterer as shown in Fig.~1.  The two dimensional Schr\"odinger
equation for the system is (we set $\hbar$=2$m$=1)

\begin{equation}
\nabla^2 \psi(x,y) = E \psi(x,y),
\end{equation}

\noindent where $E$ is the total energy of an electron and $\psi$ is
the wave function. In the leads far away from the scatterer the
electron propagates in a perfect narrow wire where eigenstates can be
decoupled into a longitudinal plane wave part $e^{ik_m x}$ and a
transverse part $\psi_m(y)$. The wave function $\psi_m(y)$
corresponds to motion between infinite potential walls along the $y$
direction is

\begin{equation}
\psi_m(y) = \sqrt{\frac{2}{b}} \sin[\frac{m\pi}{b}(y+b/2)],~~~~
m=1,2,3... \qquad.
\end{equation}

\noindent The scattering states are of the form

\begin{eqnarray}
\psi^L_{m}(x,y)={e^{ik_{m}x} \over \surd{k_{m}}}
\sin[\frac{m \pi}{b}(y+b/2)]  \nonumber +
\sum^\infty_{n=1} R_{n\,m}{e^{-ik_nx} \over \surd{k_n}}
\sin[\frac{n \pi}{b}(y+b/2)]\,\,\,\,\,for ~x \le -a,
\end{eqnarray}

\noindent and

\begin{equation}
\psi^R_{m}(x,y)=\sum^\infty_{n=1} T_{n\,m}
{e^{ik_nx} \over \surd{k_n}}
\sin[\frac{n \pi}{b}(y+b/2)] \,\,\,\,\,\, for ~x \ge a  \,\, ,
\end{equation}

\noindent where the index $n$ runs over all possible transverse
modes. The incident particle is in the $m$-$th$ mode. The
longitudinal wave vectors are given by

\begin{equation}
k_n^2=E - n^2 (\frac{\pi}{b})^2 ,
\end{equation}

\noindent $E$ being the total energy. The quantity $n^2 \pi^2/b^2$ is
the energy associated with the $n$-$th$ transverse mode and
$k_n^2/2m$ is the energy associated with the longitudinal motion of
an electron in the $n$-$th$ transverse mode.

The transmission ($T_{nm}$) and reflection ($R_{nm}$) amplitudes are
to be determined by matching solutions across $\mid x \mid = a$ and
can be expressed in terms of the matrix coefficients

\begin{equation}
R_{m\,m^\prime}=\mid {S^e_{m\,m^\prime}+S^o_{m\,m^\prime} \over 2}
\mid^2,
\end{equation}

\begin{equation}
T_{m\,m^\prime}=\mid {-S^e_{m\,m^\prime}+S^o_{m\,m^\prime} \over 2}
\mid^2,
\end{equation}

\noindent where

\begin{equation}
S^e_{m,m^\prime}=e^{-ik_ma}[1+2i(F^e-i)^{-1}]_{m,m^\prime}
e^{-ik_{m^\prime}a},
\end{equation}

\noindent and 

\begin{equation}
S^o_{m,m^\prime}=e^{-ik_ma}[1+2i(F^o-i)^{-1}]_{m,m^\prime}
e^{-ik_{m^\prime}a}.
\end{equation}

\noindent The matrices $F^e$ and $F^o$ are obtained by matching the
wave-function outside the scatterer to those inside the scatterer and
are given by (for details see Ref. \cite{bay})
\begin{equation}
F^e_{m^\prime\,m}=-{b \over c(k_{m^\prime}k_m)^{1/2}} \sum_{p=1}^\infty
I_{m^\prime\,p}I_{m\,p}K_p \tan(K_p a),
\end{equation}

\noindent and

\begin{equation}
F^o_{m^\prime\,m}=-{b \over c(k_{m^\prime}k_m)^{1/2}} \sum_{p=1}^\infty
I_{m^\prime\,p}I_{m\,p}K_p \cot(K_p a),
\end{equation}

\noindent where

\begin{equation}
I_{mn}=\frac{2}{\pi}[\frac{{\rm
sin}[(\frac{m}{b}-\frac{n}{c})\frac{\pi b}{2}]
{\rm cos}[\frac{(m-n)\pi}{2}]}{\small{m}-\frac{nb}{c}} - 
\frac{{\rm sin}[(\frac{m}{b}+\frac{n}{c})\frac{\pi b}{2}] 
{\rm cos}[\frac{(m+n)\pi}{2}]}{\small{m}+\frac{nb}{c}}] ,
\end{equation}
and $K_p$ is the wave vector along the $x$ direction associated with
the $p$-$th$ channel in a double-stub. The sums in Eqs.(9) and (10)
run over infinite values of $p$.  This is because in the interior
region of the scatterer, infinite number of states (one from each
transverse mode) couple to the electronic state of incident energy
$E$. Some of them are propagating modes and some of them are
evanescent modes. The sums in Eqs.(9) and (10) are highly converging
and one can truncate them appropriately depending on the accuracy
required. Typically 10 evanescent modes are included in our
calculations.

Having found the complex transmission and reflection amplitudes
across one such scatterer we use the scattering matrix cascading
technique to find the transmission across many such scatterers put
together in series. The separation between the two scatterers is
$l$ and all other lengths $b$, $a$, $c$ associated with the
scatterer are the same as in Fig.~1. For details of the method of
cascading we refer to Ref. \cite{len}. Here we just outline the
algorithm.

The scattering matrix (S) for any symmetric scatterer is given by

\begin{equation}
\pmatrix{{r} & {t^\prime}\cr
{t} & {r^\prime}\cr},
\end{equation}

\noindent where the $r$ and $t$ are matrices:

\begin{equation}
r=\pmatrix{{R_{11}} & {R_{12}} & \ldots & {R_{1n}}\cr
\vdots & \vdots & \ddots & \vdots \cr
{R_{n1}} & {R_{n2}} & \ldots & {R_{nn}}\cr},
\end{equation}

\noindent and

\begin{equation}
t=\pmatrix{{T_{11}} & {T_{12}} & \ldots & {T_{1n}}\cr
\vdots & \vdots & \ddots & \vdots \cr
{T_{n1}} & {T_{n2}} & \ldots & {T_{nn}}\cr}.
\end{equation}

For the case of symmetric scatterers considered in our work we have
$t^\prime=t$ and $r^\prime=r$.  There are $n$ incoming channels and
$n$ outgoing channels. If two scatterers are characterized by the S
matrices $S_1$ and $S_2$ then the resultant scattering matrix, on
cascading the two systems, is the direct product of $S_1$ and $S_2$.
The total S matrix of N scatterers is the direct product of N
scattering matrices. The order in the direct product of the N
S-matrices must be kept the same as that of the spatial order of the
scatterers. This is because the S matrices do not commute. Thus from
the total S matrix we can find the two terminal-conductance G, which
is given by the Landauer formula \cite{lan} {\em i.e.}, G=${2e^2\over
h}\sum_{ij} \mid T_{ij} \mid^2$.

\section{Results and discussions}

We first consider the system of a quantum channel (main wire) in
which a series of double-constrictions are arranged periodically.
The system under consideration consists of 87 double-constrictions.
We have taken $a/b=0.5$, $c/b=0.805$. The distance between two
consecutive constrictions is $l/b=0.5$ where $b$ is the width of the
main wire along the $y$ direction. Here, $2a$ and $c$ are the widths
of a single constriction along the $x$ and $y$ direction,
respectively. The dimensionless conductance, g ($= \frac{h}{2e^2}G$)
for this system is plotted in Fig.~2 (dotted curve) as a function of
dimensionless incident energy $Eb_{1}^2$ (where $b_{1}=b$/2 is the
half width of the main wire). We vary the incident energy in the
range such that two to seven propagating modes in the main wire are
occupied. We find a conductance quantization which is not
monotonically increasing but the quantization is not good. At higher
energies, since there are a lot of propagating modes, there are a lot
of nearly degenerate channels and small mode mixing can wash out the
quantization. And at lower energies, in the case of such a large
value of $2a$ ($2a/l$=1), there are not many Fano resonances in the
system to produce a good quantization. The reason is that the
exponentially decaying evanescent modes have a negligible probability
density inside the constriction. A breakdown of the symmetry by the
presence of a defect can completely destroy the quantization as
demonstrated in Fig.~2 (solid curve).  We replace the middle
constriction of the original system by a different constriction.  The
width of the newly placed constriction (defect constriction) is such
that $c/b=0.905$. Such a defect in a QPC is a non-adiabatic or a very
strong scatterer. Other parameters are the same as for the other
constrictions. Fig.~2 (solid curve) shows the dimensionless
conductance ($g$+3) as a function of the dimensionless incident
energy for this system. The incident energy is varied to excite two
to seven modes in the main wire. We see that the quantization of the
conductance is lost except for some of the lowest steps of the few
lowest sub-bands.  Thus a single defect is enough to destroy the
quantization of the conductance of the system.

We can increase the probability density of the modes in the
scatterer, {\it i.e.,} by taking a double-stub or a very small
double-constriction.  In case of the double-stub there will be more
propagating modes in the scatterer and in case of the short
double-constriction the evanescent modes in the constriction will
have a large probability density. In the following we demonstrate
that in these two cases we get very good quantization and also they
are robust even in the presence of the defect.

First we consider a system comprising of a series of identical
double-stubs arranged periodically in a point contact. We take 87
double-stubs. Other parameters for this system are given by $a/b=0.5$
and $c/b=1.055$. The distance between two consecutive double-stubs is
$l/b=0.5$. Here, $2a$ and $c$ are the widths of a single double-stub
along the $x$ and $y$ directions, respectively.  Fig.~3 (dotted
curve) shows the dimensionless conductance for this system as a
function of dimensionless incident energy. In this case we see that
the quantization is very good.  So, we see that a system consisting
of a series of double-stubs in a point contact gives better
conductance quantization than that of a series of constrictions. This
is in contrast to the fact that a single double-stub in a QPC
destroys quantization more effectively than a single constriction
\cite{sza}.  We now put a defect in the system.  The defect is
introduced by the replacement of the middle double-stub in the system
by another one whose width in the $y$ direction, taken to be $c/b$ =
1.205, is different from the other double-stubs.  We excite up to
seven modes in the main wire by a proper choice of the range of
incident energy.  The dimensionless conductance ($g$+3) for this
system (the system with a defect) is plotted as a function of
dimensionless incident energy.  This is shown in Fig.~3 (solid curve)
where we notice that the quantization of the conductance still partly
survive and is not destroyed completely.  Quantization is still
retained up to the 4th or 5th step.  Backscattering in some regimes
is negligibly small as found by comparing the solid curve with the
dotted curve in Fig.~3.

Next, in Fig.~4 (dotted curve) we consider a system that consists of
87 small double-constrictions regularly placed in a main wire. We
have taken $a/b=0.125$ (so it is a small constriction), $c/b=0.805$,
and the distance between two consecutive constrictions is $l/b=0.5$.
Note that the only difference in this case, {\it i.e.,} the value of
$a$, is very small compared to the system considered in Fig.~2.
Again we restrict ourselves to maximum seven mode propagation.  Here
we see that the quantization of the conductance is much better than
that in Fig.~2. We have examined the quality of quantization of
conductance for systems with various values of $a/b$ and noticed that
the quality of the quantization of conductance degrades as the the
value of $a/b$ deviates from 0.125 such that $a/b$ and $l/b$ are
incommensurate.  Now we introduce a defect constriction in the system
as just discussed in the above in order to simulate mode mixing and
backscattering.  We replace the middle constriction by a defect
constriction. The value of $c/b$ for the defect constriction is 0.905
while the other parameters are the same as for the other
constrictions.  The dimensionless conductance ($g$+3) of the system
is plotted as a function of dimensionless incident energy in Fig.~4
(solid curve).  Two to seven propagating modes are considered.  We
notice that the quantization of conductance for the lower bands are
fairly retained even in the presence of this defect.

\section{Conclusions}

The quantization of conductance in periodic channels is more robust
than that in un-modulated point contacts. The experiment of
Kouwenhoven {\it et al} \cite{kou} measures the conductance of a
periodic channel but one has to take a longer chain to see the
quantization.  But because of the robustness of the effect it will be
worthwhile to do such an experiment in a long channel. A system of
periodic stubs is more robust than that of a system of periodic
constrictions. But if the spacing between the constrictions is made
very small, then the quantization of periodic constrictions is
equally robust. The Fano resonances, that are always present in a
multichannel scattering problem, have a strong influence on the
quantization of the conductance.

\section{Acknowledgments}

One of us (PSD) thanks Prof. N. Kumar for useful discussions.  This
work is supported by the Flemish Science Foundation(FWO-Vl) grant No:
G.0277.95, and the Belgian Inter-University Attraction Poles
(IUAP-VI).  One of us (PSD) is supported by a scholarship from the
University of Antwerp and FMP is a Research Director with FWO-Vl.

\centerline{\bf Figure captions}

\noindent Fig.1.  Schematic diagram of a Quantum point contact with
a non-adiabatic scatterer in it. The axes directions and the various
length parameters are shown in the figure.

\noindent Fig.2.  (dotted curve) Plot of dimensionless conductance
(g) versus dimensionless Fermi energy ($Eb_1^2$) for a system of 87
double-constrictions ($a/b=0.5,\,c/b=0.805,\,l/b=0.5$).  (solid
curve) Plot of dimensionless conductance (g) versus dimensionless
Fermi energy ($Eb_1^2$) for a system that has a defect at the center
of the system considered in the dotted curve. For the central
double-constriction $c/b=0.905$.

\noindent Fig.3.  (dotted curve) Plot of dimensionless conductance
(g) versus dimensionless Fermi energy ($Eb_1^2$) for a system of 87
double-stubs ($a/b=0.5,\,c/b=1.055,\,l/b=0.5)$.  (solid curve) Plot
of dimensionless conductance (g) versus dimensionless Fermi energy
($Eb_1^2$) for a system that has a defect at the center of the system
considered in the dotted curve. For the central double-constriction
$c/b=1.205$.

\noindent Fig.4. (dotted curve) Plot of dimensionless conductance (g)
versus dimensionless Fermi energy ($Eb_1^2$) for a system of 87
double-constrictions ($a/b=0.125,\,c/b=0.805,\,l/b=0.5)$.  (solid
curve) Plot of dimensionless conductance (g) versus dimensionless
Fermi energy ($Eb_1^2$) for a system that has a defect at the center
of the system considered in the dotted curve. For the central
double-constriction $c/b=0.905$.


\begin{thebibliography}{99}
\bibitem[*]{eml1} email address: deo@uia.ua.ac.be
\bibitem[o]{eml2} email address: peeters@uia.ua.ac.be
\bibitem{wee} B. J. van Wees, H. van Houten, C. W. J. Beenakker,
J. G. Williamson, L. P. Kouwenhoven, D. van der Marel, and C. T.
Foxon, Phys. Rev. Lett. {\bf 60}, 848 (1988).
\bibitem{wha} D. A. Wharam, T. J. Thornton, R. Newbury, M. Pepper,
H. Ahmed, J. E. Frost, D. G. Hasko, D. C. Peacock, D. A.
Ritchie, and G. A. C. Jones, J. Phys. C {\bf 21}, L209 (1988).
\bibitem{lan} R. Landauer, Z. Phys. B {\bf 68}, 217 (1978).
\bibitem{tap} T. Chakraborty and P. Piettilienen, {\it The
Fractional Quantum Hall Effects}, Berlin; Springer Verlag (1988).
\bibitem{bag} P. F. Bagwell, Phys. Rev. B, {\bf 41}, 10354 (1990).
\bibitem{ber} K. F. Bergrenn and Z. Ji, Phys. Rev. B, {\bf 43},
4760 (1991); $ibid$. {\bf 45}, 6650 (1992).
\bibitem{kir} G. Kirczenow, Solid State Commun. {\bf 68}, 715 (1988).
\bibitem{sza} A. Szafer and A. D. Stone, Phys. Rev. Lett. {\bf
62}, 300 (1989).
\bibitem{hua} Hua Wu, D. W. L. Sprung, J. Martorell and S.
Klarsfeld, Phys. Rev. B {\bf 44}, 6351 (1991).
\bibitem{cir} E. Tekman and S. Ciraci, Phys. Rev. B. {\bf 43},
7145 (1991) and references therein.
\bibitem{pee} F. M. Peeters, {\it Science and Engineering of one and zero
Dimensional Semiconductors}, Vol. 214 of NATO Advanced Study Institute
Series B Physics, eds. S. P. Beaumont and C. M. Sotomayor-Torres
(Plenum Press, New York, 1990), p.107.
\bibitem{nix} A. Nixon, J. H. Davies, and H. U. Baranger, Phys. Rev.
B. {\bf 43}, 12638 (1991).
\bibitem{but} P. N. Butcher and J. A. McInnes, J. Phys. Cond.
Mat. {\bf 7}, 745 (1995).
\bibitem{kou} L. P. Kouwenhoven, F. W. J. Hekking, B. J. van Wess, C. J.
P. M. Harmans, C. E. Timmering, and C. T. Foxon, Phys. Rev. Lett. 
{\bf 65}, 361 (1990).  
\bibitem{bru} J. A. Brum, in Proc. of 20th International
conference on {\it The Physics of Semiconductors}, Vol.3, 
ed. E. M. Anastassakis and J. D. Joannopoulos, World 
Scientific, Singapore, p.2363.
\bibitem{aki} R. Akis, P. Vasilopoulos and P. Debray, Phys. Rev.
B {\bf 52}, 2805 (1995).
\bibitem{noc} J. U. N$\ddot o$ckel, Phys. Rev. B {\bf 45},
14225 (1992).
\bibitem{len} M. Leng and C. S. Lent, Phys. Rev. Lett. {\bf 71} 137
(1994); M. Leng and C. S. Lent, Phys. Rev. B {\bf 50}, 10832(1994).
\bibitem{pro} P. Singha Deo and A. M. Jayannavar, Physica B {\bf 228},
353 1996. 
\bibitem{bay} B. F. Bayman and C. J. Mehoke, Am. J. Phys.
{\bf 51}, 875 (1983).

\end{thebibliography}
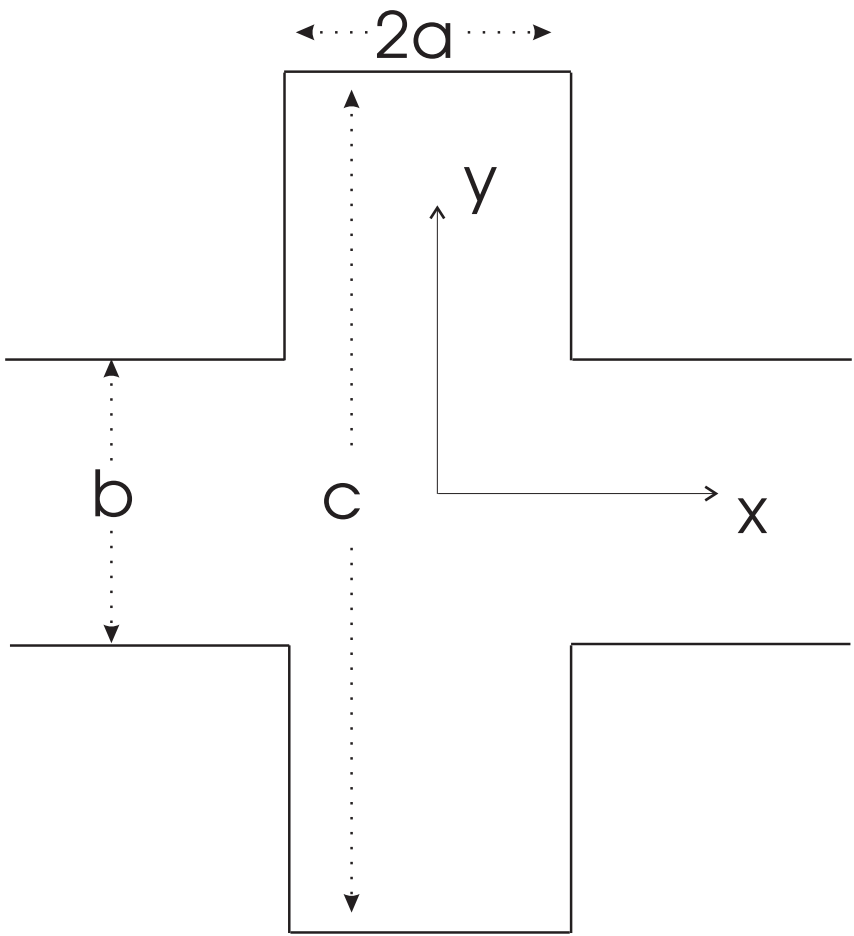
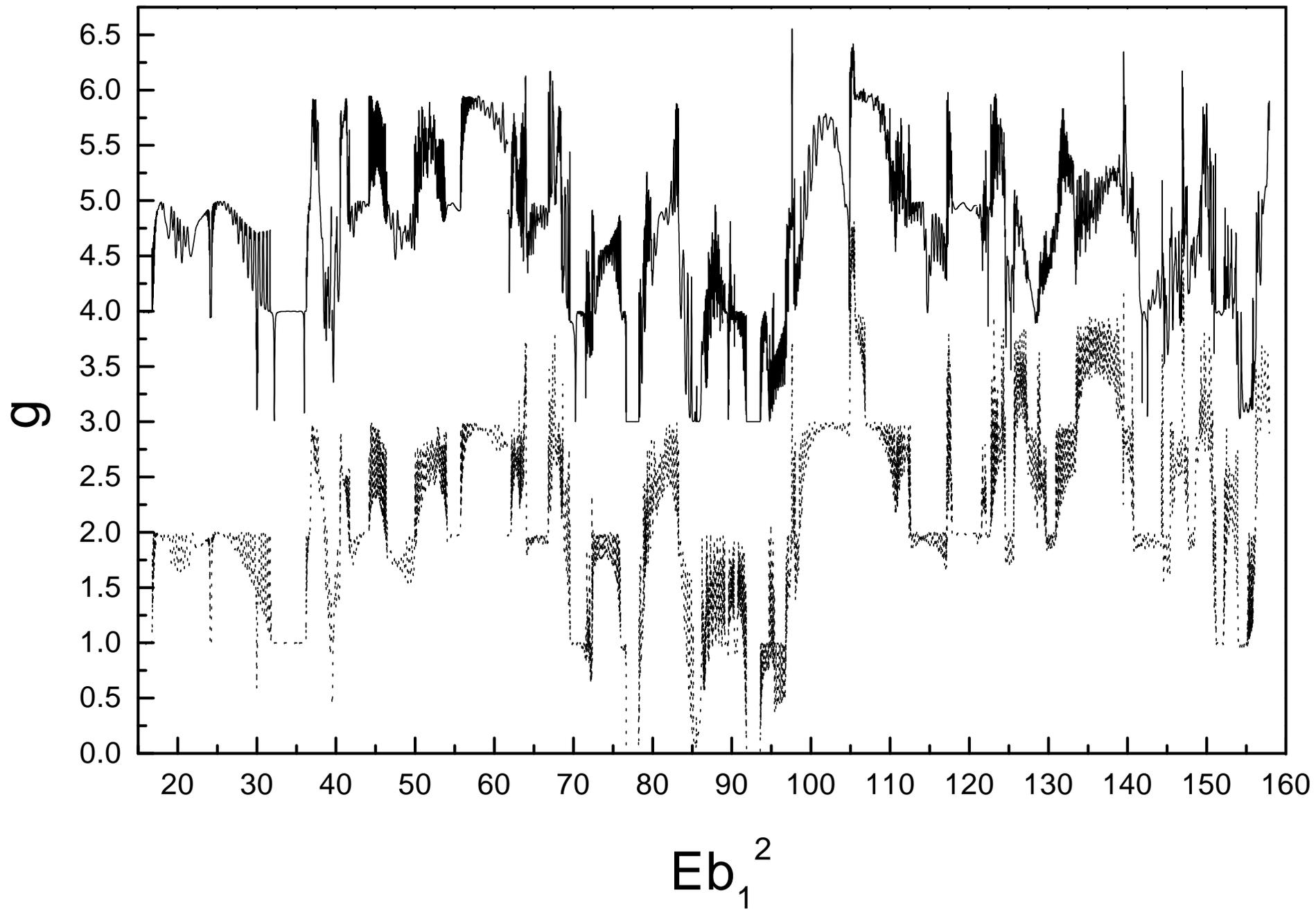
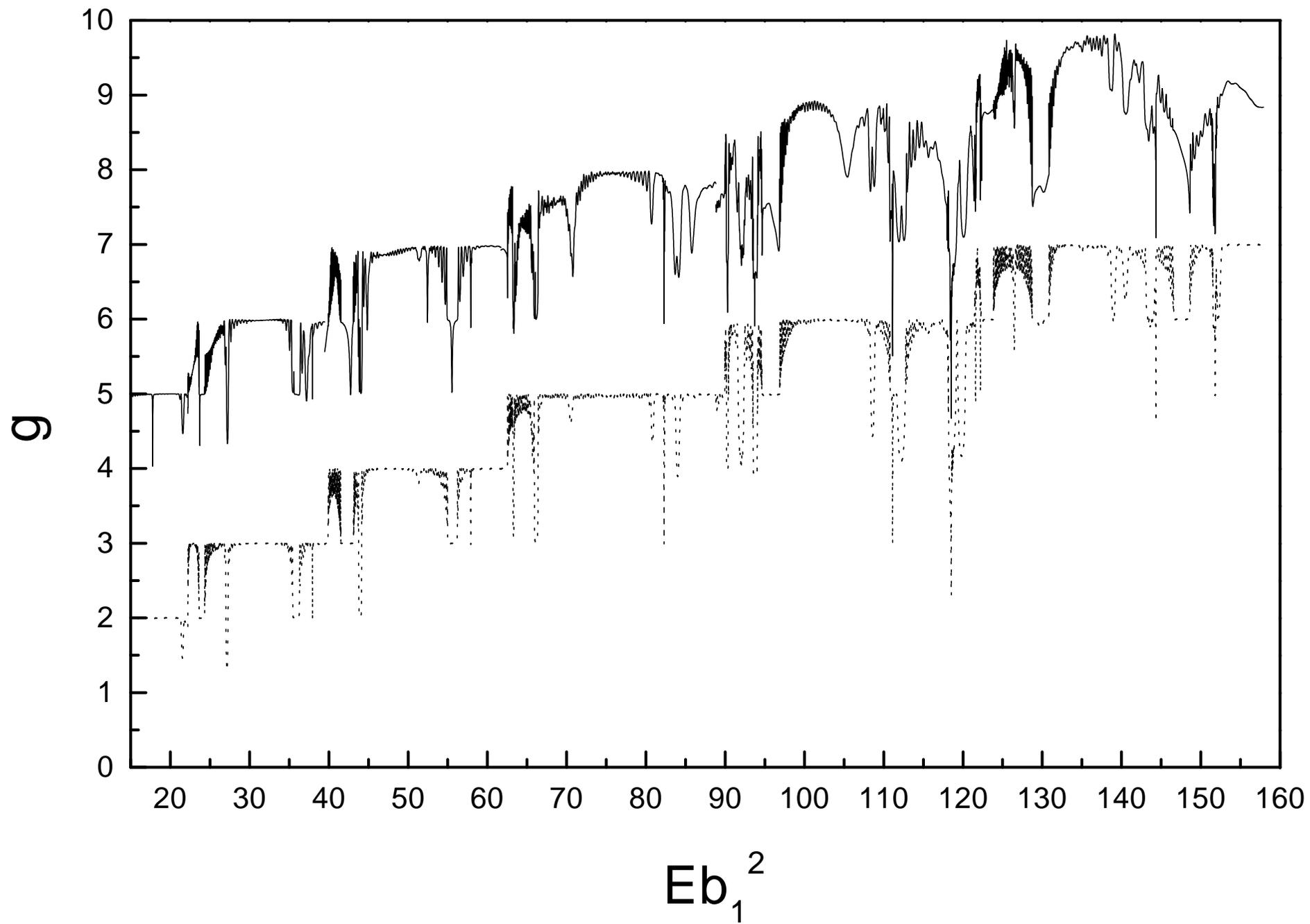
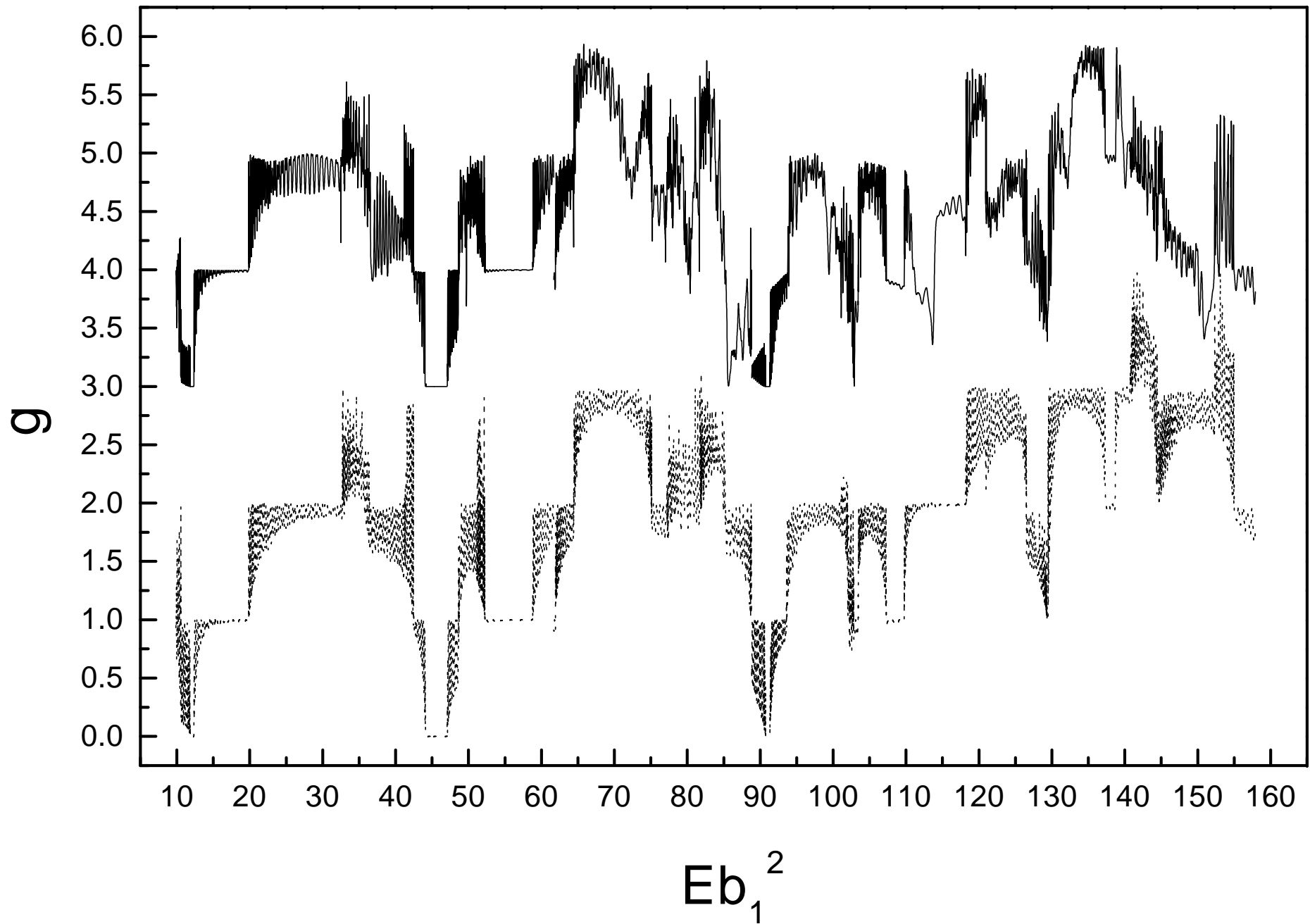
\end{document}